%% file: main.tex
\documentclass[screen, acmsmall, table]{acmart}

\AtBeginDocument{%
  }

\input{meta/packages}


\begin{document}

\input{meta/authortitle}

\input{sections/00_abstract}

\input{meta/acm}

\maketitle

\input{sections/01_introduction}
\input{sections/02_related_work}

\input{sections/03_pre_study}
\input{sections/04_main_study}
\input{sections/05_result}
\input{sections/06_discussion}

\input{sections/07_conclusion}

\bibliographystyle{ACM-Reference-Format}
\bibliography{ref}
\appendix
\input{sections/09_appendix}

\end{document}

%% file: meta/packages.tex
\usepackage{graphics} 
\usepackage{dirtytalk}
\usepackage{adjustbox} 
\usepackage{color}
\usepackage{booktabs}
\usepackage{textcomp}
\usepackage{multirow}
\usepackage{paralist}
\usepackage{array}
\usepackage{wrapfig}
\usepackage{todonotes}
\usepackage{diagbox}
\usepackage{makecell}
\usepackage{subfigure}
\usepackage{tablefootnote}
\usepackage[nolist]{acronym}
\usepackage{caption}
\usepackage{float}

\begin{acronym}
    \acro{HCI}{Human-Computer Interaction}
    \acro{CSCW}{Computer-Supported Collaborative Work}
    \acro{CS}{Computer Science}
    \acro{MSE}{Mean Square Error}
    \acro{MWU}{Mann-Whitney U}
    \acro{LIWC}{Linguistic Inquiry and Word Count}
    \acro{PANAS}{Positive And Negative Affect Schedule}
\end{acronym}

\newcommand\rev[1]{\textcolor{black}{#1}}
\newcommand\mrev[1]{\textcolor{black}{#1}}

\graphicspath{{figures/}}
\definecolor{tg1}{RGB}{230,195,215}  %
\definecolor{tg2}{RGB}{193,228,252}  %

%% file: meta/authortitle.tex
\title{Emotionally Aware Moderation: The Potential of Emotion Monitoring in Shaping Healthier Social Media Conversations}

\author{Xiaotian Su}
\email{xiaotian.su@inf.ethz.ch}
\orcid{0009-0004-0548-1576}
\affiliation{%
  \institution{ETH Zurich}
  \city{Zurich}
  \country{Switzerland}
}

\author{Naim Zierau}
\email{naim.zierau@unisg.ch}
\orcid{0000-0002-9451-8577}
\affiliation{%
  \institution{University of St.Gallen}
  \city{St. Gallen}
  \country{Switzerland}
}

\author{Soomin Kim}
\email{smsoominkim@gmail.com}
\orcid{0000-0003-2523-7808}
\affiliation{
  \institution{Seoul National University}
  \city{Seoul}
  \country{South Korea}
}

\author{April Yi Wang}
\email{april.wang@inf.ethz.ch}
\orcid{0000-0001-8724-4662}
\affiliation{%
  \institution{ETH Zurich}
  \city{Zurich}
  \country{Switzerland}
}

\author{Thiemo Wambsganss}
\email{thiemo.wambsganss@bfh.ch}
\orcid{0000-0002-7440-9357}
\affiliation{%
  \institution{Bern University of Applied Sciences}
  \city{Bern}
  \country{Switzerland}
}

\renewcommand{\shortauthors}{Su et al.}
\renewcommand{\shorttitle}{Emotionally Aware Moderation}

%% file: sections/00_abstract.tex
\begin{abstract}
Social media platforms increasingly employ proactive moderation techniques, such as detecting and curbing toxic and uncivil comments, to prevent the spread of harmful content. 
Despite these efforts, such approaches are often criticized for creating a climate of censorship and failing to address the underlying causes of uncivil behavior. 
Our work makes both theoretical and practical contributions by proposing and evaluating two types of emotion monitoring dashboards to users' emotional awareness and mitigate hate speech. 
In a study involving 211 participants, we evaluate the effects of the two mechanisms on user commenting behavior and emotional experiences. The results reveal that these interventions effectively increase users' awareness of their emotional states and reduce hate speech. However, our findings also indicate potential unintended effects, including increased expression of negative emotions (Angry, Fear, and Sad) when discussing sensitive issues. These insights provide a basis for further research on integrating proactive emotion regulation tools into social media platforms to foster healthier digital interactions.
\end{abstract}

%% file: meta/acm.tex
\ccsdesc[500]{Human-centered computing~Human computer interaction (HCI)}
\ccsdesc[300]{Computing methodologies~Natural language processing}
\ccsdesc[100]{Empirical investigations}

\keywords{social media, proactive intervention, emotion monitoring}

%% file: sections/01_introduction.tex
\section{Introduction}
Social media provides unprecedented freedom for communication and self-expression. However, this openness is frequently exploited to disseminate violent messages, derogatory comments, and hateful speech \cite {cscw11-quality-discourse}. This phenomenon, known as online hate speech, is defined as a \say{direct and serious attack on any protected category of people based on their race, ethnicity, national origin, religion, sex, gender, sexual orientation, disability or disease} \cite{cscw20-hate, aaai18-hatelingo}.

Hate speech is increasing at an alarming rate across social media platforms \cite{SHETH2022312, cyber21, cscw11-quality-discourse}. 
This increase not only leads to a deteriorated environment on the internet but can also translate into violence outside the virtual world \cite{physical11}. For instance, a study reported that between 15\% and 35\% of young people have been victims of cyberbullying \cite{rout10-bullying}. Following these incidents, two-thirds of teens felt a decline in self-esteem, nearly a third reported affected friendships, and 13\% indicated a negative impact on their physical health \cite{24security_cyberbullying}.
These numbers illustrate the devastating effect of harassing online messages on many people's well-being and the social fabric overall, highlighting the urgency to foster more inclusive and constructive dialogue on social media platforms. 

Most existing interventions for addressing harmful content on social media adopt a reactive approach, focusing on removing offensive content after it has been posted. This is typically achieved through either automated systems \cite{VALLECANO2023119446, hatebert21} or manual moderation \cite{cscw23-convex, cscw19-crossmod}. However, reactive methods are  criticized for both their potential to suppress free expression \cite{cscw21-censorship} and for their delayed intervention, addressing harmful content only after it appears online \cite{kiesler2012regulating}.Furthermore, such methods often lack the nuance required to address the underlying emotional dynamics of online interactions. Uncivil behavior often originates from regular users caught in highly charged or tense situations, exacerbated by the social and technological affordances of social media platforms \cite{cscw22-thread}.

In contrast, interventions like those designed by \citet{cscw19-gremobot} aim to proactively regulate group emotions, following theories of emotion regulation. However, these approaches still followed a reactive paradigm, intervening only after the content was posted. There are some proactive interventions designed to promote constructive dialogue on social media, encouraging users to reflect on their messages before posting  (e.g., \cite{cscw19-accountability, CAPTCHAs19, royen2022think}). While promising, these methods are generally static and applied uniformly, failing to account for the diverse communication styles and emotional responses of individual users \cite{KAZIENKO202343}. Recent studies by \citet{cscw21-recast} and \citet{cscw22-thread} have explored personalized, real-time support to reduce toxic communication. However, these approaches rely solely on machine learning to analyze text, overlooking users' emotional states. 

In contrast to these existing approaches, our emotion regulation intervention follows a proactive, personalized approach by designing two emotion intervention mechanisms and developing a specialized dashboard. Specifically, we aim to provide insight into the emotional tone of social media posts and assist users with emotion regulation before they post. This dashboard distinguishes between two types of emotion monitoring interventions: self monitoring and peer monitoring. Self monitoring enhances self-awareness and emotional regulation by reflecting the user's own emotions, while peer monitoring enhances peer-awareness and social understanding by reflecting the emotions of others. This study evaluates the benefits and potential pitfalls of each monitoring type, focusing on their impact on users' emotional awareness, social interactions, and the reduction of hate speech. Specifically, we aim to answer the following research questions:

\begin{enumerate}
  \item How does real-time emotion \emph{peer monitoring} influence users' messages in online discussions compared to environments without such monitoring?
  \item How does real-time emotion \emph{self monitoring} influence users' messages in online discussions compared to environments without such monitoring?
\end{enumerate}

Within the context of our main study, we examined whether our two key interventions—providing feedback on (1) the peer's and (2) users' emotional sentiments—can impact user commenting behaviors and emotional experiences using both quantitative and qualitative data. While the effects noted in our main study were modest and limited by its scale, they, combined with our qualitative insights, indicate enhancing users' emotional awareness could influence social media interactions in various ways. This approach has the potential to contribute positively to existing strategies for promoting healthier, more constructive interactions, but it may also have unforeseen negative impacts on users' emotional experiences as our study shows. This research lays a foundation for future explorations, offering specific avenues for integrating natural language processing-based user-centered tools into social media platforms to make social media discussions more positive and productive.
In this research, we present three contributions: 
\begin{itemize} 
 \item We introduce a mechanism designed to enhance users' emotional awareness and reduce toxic speech in social media conversations.
 \item We develop a pair of user interventions, manifested as an emotion monitoring dashboard, which operationalizes this mechanism within social media platforms.
 \item We provide evidence from a comprehensive user study with a total of 211 participants to \mrev{investigate the effect} of our intervention. The studies offer preliminary findings on the multi-faceted impact of our intervention, demonstrating their influence on emotional awareness and users' commenting behavior.
\end{itemize}

%% file: sections/02_related_work.tex
\section{Related Work}
Our work builds on the extensive research on online discussion moderation.
\rev{In this section, we first survey the types of existing content moderation practices along the reactive-proactive spectrum, characterizing their effectiveness and costs.}
\begin{figure*}
    \centering
    \includegraphics[width=0.9\columnwidth]{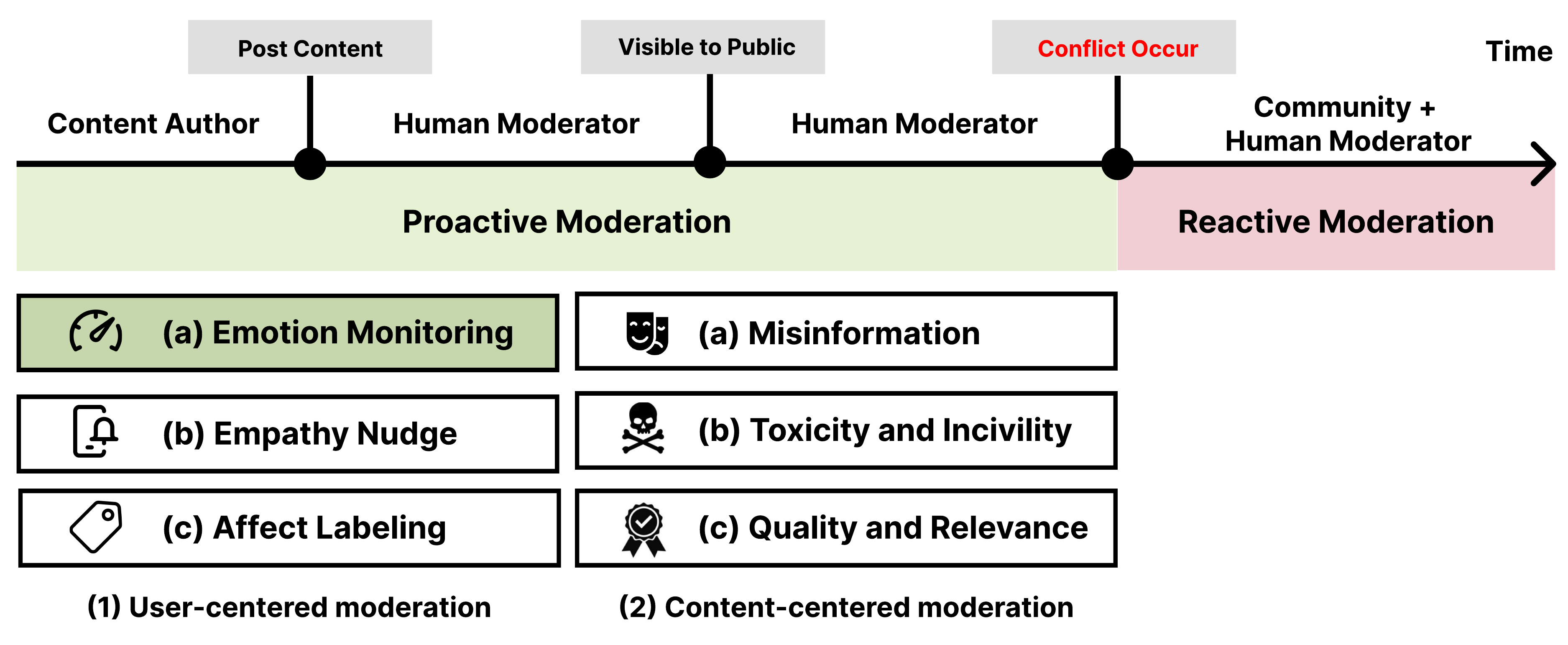}
    \caption{\rev{Our work (Emotion Monitoring) builds on existing text-based moderation practices on social media along the proactive-reactive spectrum. Our emotion monitoring dashboard aims to enhance users' emotional awareness and help them construct more neutral messages.}}
    \label{fig:rw}
\end{figure*}
\subsection{\rev{Content Moderation Strategies on Social Media}}
The escalating spread of harmful material on digital platforms presents a significant societal challenge, encompassing various forms like hate speech, offensive language, bullying and harassment, misinformation, spam, violence, graphic content, sexual abuse, self-harm, and many others. Researchers in the \ac{HCI} and \ac{CSCW} fields have a growing interest in content moderation to foster more positive and constructive online discourse \cite{cscw22-build, cues92, sproull1986reducing}. An emerging line of work proposes distinguishing different moderation practices based on the timing of the intervention relative to the objectionable action. 

As shown in Figure \ref{fig:rw}, proactive moderation is designed to prevent such behavior from occurring in the first place \cite{cscw22-thread, cscw22-proactive}, either remind the content author to reflect on their texts before posting or detect and filter out the inappropriate content by human moderators after users posting. Another approach is reactive moderation \cite{seering2019moderator}, where people respond to conflicts or negative signals after they occur. This approach utilizes a combination of community involvement and human moderation to address problematic content. The community plays a role in identifying and flagging problematic content through collaborative efforts, such as users on Quora downvoting an unfriendly comment. At the same time, human moderators can intervene and take appropriate action against both the sender and recipients of the content. For example, a moderator may remove the unfriendly comment and hide it from the user who received it. Various tools have been developed to support moderators' tasks across different platforms \cite{cscw23-convex, cscw19-crossmod}.

Reactive moderation has inherent weaknesses while widely used and studied \cite{cscw23-convex, cscw19-transparency, cscw19-crossmod}. 
Firstly, it involves taking action against uncivil content that has already been posted, allowing the offending content to be seen and potentially spread before moderators can intervene \cite{kiesler2012regulating}. 
This can harm platforms and their users and can also adversely affect moderators who must regularly view and take action on disturbing content, threatening their emotional well-being \cite{cscw23-moderation, cscw21-punishment} and even their personal safety \cite{chi19-moderator}. 
These harms underscore the need to explore alternative moderation approaches that can mitigate such human suffering.

Proactive approaches to moderation aim to discourage undesirable actions or to encourage prosocial behavior and productive conversations \cite{cscw17-proactive}. 
Some studies explored indirect methods, like prompting users with tasks before commenting to prime positive engagement (e.g., completing a captcha \cite{CAPTCHAs19}). 
For more direct interventions, \citet{cscw21-recast} reports the toxicity score of users' textual input, providing immediate feedback on potentially harmful content. 
This is argued as a more effective way to protect online communities from harm because it reduces the amount of uncivil content that gets created in the first place \cite{kiesler2012regulating}.

\subsection{Algorithmic Approaches for Proactive Moderation}
Many algorithmic approaches have been developed for proactive content moderation. \rev{These methods aim to validate the authenticity and quality of public content, addressing critical issues such as misinformation and toxicity (Figure \ref{fig:rw}.2).
For instance, \citet{WSDM18-leveraging} developed a scalable online algorithm to reduce the spread of fake news and misinformation. \citet{risch2020toxic} presented various deep learning approaches for detecting toxic comments while \citet{ieee23-fakenews} proposed a methodology to identify toxic counterfeit news articles across social media platforms. Complementing these efforts, \citet{www09-learning} developed a semi-supervised coupled mutual reinforcement framework that simultaneously evaluates content quality and user reputation.}

\rev{Despite the effectiveness of proactive approaches in detecting toxicity \cite{cscw21-recast} and incivility \cite{cscw22-thread} in certain contexts, these methods lack personalized guidance for user commenting behavior \cite{KAZIENKO202343}. And they can sometimes exacerbate tensions or make users feel unfairly judged or censored \cite{cscw21-censorship}. }
Such approaches often lack the nuance required to address the underlying emotional dynamics of online interactions, resulting in a one-size-fits-all strategy that overlooks individual differences in communication styles and emotional responses \cite{KAZIENKO202343}. 
\rev{To address these limitations, we propose integrating emotional awareness, a critical factor in fostering prosocial behavior \cite{lockwood2014emotion, empathy11}, into content moderation mechanisms. Emotional awareness encompasses both the ability to recognize one’s own emotions and the capacity to understand them in context \cite{van2003continuing}. Extensive research underscores the role of emotional capacities, such as emotion regulation, in enabling individuals to empathize with others and respond to suffering without becoming overwhelmed \cite{geangu2011individual, henschel2020emotion}. Building on this understanding, we advocate for a user-centered moderation approach (Figure \ref{fig:rw}.1a) that prioritizes users' emotional dynamics over traditional content-centered moderation strategies (Figure \ref{fig:rw}.2).}

\subsection{\rev{Text-based Emotion Regulation on Social Media}}
\rev{
The ability to regulate emotions is essential across nearly every domain of life \cite{gross1998emerging}. There is a growing interest in \ac{HCI} to envision, design, and evaluate technology-enabled interventions that support users’ emotion regulation \cite{chi23-designing, chi22-digital}.
Recognizing and managing emotions is particularly challenging in online text-based communication \cite{chi21-applying}. The lack of nonverbal cues in online communication deteriorates the ability to control emotions and empathize with others \cite{93impression}. 
To address these challenges, researchers have investigated various user-centered approaches (Fig \ref{fig:rw}.1) to encourage emotion regulation and foster emotional awareness in online discussions. 
For example, \cite{royen2022think} explored designs to support emotion regulation by encouraging reflective behaviors, while \cite{cscw19-accountability} explored how different types of empathy nudges can promote prosocial behaviors (Figure \ref{fig:rw}.1b).
\citet{saldias2019tweet} reported that users produced a significantly higher percentage of neutral content when had access to emotional statistics about their posts. \citet{mindterk21-exploration} explored how labeling emotions in user interfaces for online news articles can assist in emotion regulation (Figure \ref{fig:rw}.1c). Collectively, these findings highlight the promise of integrating emotional awareness into the design of online platforms as a means of fostering healthier, more supportive digital communities.
However, current approaches are predominantly static and do not adapt to the dynamic nature of online interactions or the unique needs of individual users. To effectively enhance interpersonal communication and emotional well-being on social media, it is crucial to implement adaptive emotion regulation strategies that can swiftly respond to the diverse emotional expressions and communication styles of users \cite{chi23-designing}. To address this challenge, we propose the development of an emotion monitoring dashboard—a novel tool designed to boost emotional awareness and minimize offensive comments by providing users with real-time feedback on the emotional tone of their own and others' messages.
}

\rev{Although emotion regulation is a well-established concept in HCI, there is limited research on emotion regulation interventions in online discussions, with the GremoBot project being a notable exception \cite{cscw19-gremobot}. We build on the framework proposed by Slovak et al. \cite{chi23-designing} to understand this work's unique contributions regarding (1) the target of emotion regulation initiatives, (2) their delivery mechanisms, and (3) their design components.
First, GremoBot aimed to regulate group behavior in work discussion settings, whereas our study focuses on providing emotion feedback to individual users in social media discussions. Second, while both studies employ a didactic approach with a recommender system to facilitate emotional awareness, GremoBot provides feedback reactively—after a prolonged pause in discussion or when a certain level of negativity is detected. In contrast, our study adopts a proactive approach, offering feedback before a comment is posted. Third, GremoBot delivers analyses and recommendations at the group level, with all participants receiving the same suggestions. In our study, we provide personalized analyses and recommendations tailored to individual users.
Taken together, our work marks a significant shift from reactive group-level correction to proactive, preemptive individual-level support. By enabling users to regulate their emotions before posting potentially toxic content, we offer personalized feedback and actionable recommendations. By addressing the emotional and social complexities of online interactions, our study aims to go beyond merely moderating behavior. Instead, we seek to foster genuine understanding and connection among users, paving the way for healthier and more empathetic digital environments. This approach allows us to investigate the timing and effectiveness of these strategies in real-time, offering practical design guidance for emotion regulation tools.
Moreover, while the GremoBot project featured an early-stage exploration of an innovative design space with 27 participants, our study advances the field by conducting a large-scale experimental evaluation involving 211 participants. This responds to recent calls for large-scale experimental studies and randomized controlled trials to provide causal evidence on the effectiveness of technology-facilitated emotion regulation interventions \cite{chi23-designing}}.

%% file: sections/03_pre_study.tex
\section{\rev{Overview of The Methodology}}
\input{tables/demographics}
\rev{As shown in Table \ref{tab:demographics}, our research encompassed two preliminary studies  (Section \ref{sec:preb}) and one comprehensive main study (Section \ref{sec:main}) to rigorously test our hypotheses. In the first pre-study, we explored various topics to select an emotionally engaging subject as the foundation. The second pre-study then aimed to confirm our premise that users' level of emotional awareness can be raised in emotionally charged social media interactions. Culminating in the main study, we applied these interventions, specifically emotion peer and self monitoring tools, in a simulated online discussion on abortion. By uncovering qualitative and quantitative insights into participant engagement with and reactions to the emotion monitoring feature, we provide a comprehensive evaluation of our hypotheses in a controlled, yet realistic, social media setting.}

\section{Pre-studies}
\label{sec:preb}
Before our main study, we conducted two pre-studies focusing on the topic involvement and emotional awareness. The topic involvement study aims to identify a topic that not only resonates emotionally with social media users but also frequently encompasses hate speech or toxic comments. The emotional awareness study aims to provide initial insights into the potential of increased emotional awareness as a strategy to mitigate toxic communication. The details on the demographics of participants can be found in Table \ref{tab:demographics}.

\subsection{Topic involvement}
\input{tables/prea}
This study aims to identify a topic that not only resonates emotionally with social media users but also frequently encompasses hate speech or toxic comments. \rev{Research has examined controversial topics on social media around policy change and activism, such as abortion \cite{zhang2016gender}, climate change \cite{segerberg2011social}, and racial inequality \cite{de2016social}.}
In collaboration with a senior researcher \rev{in Cognitive Science}, we curated three topics known to provoke toxic communication on social media: abortion, the role of television, and vegetarianism. 
To determine the most emotionally engaging topic, we enlisted \mrev{41} participants from Prolific (M (age) = 36.9, SD (age) = 12.6; 46.3\% male, 48.8\% female, 4.9\% other).
Each participant engaged in separate social media discussions on these three topics, presented in randomized order. 
Each discussion scenario consisted of a newspaper article and a user comment, which they were asked to write a response to. After each scenario, \rev{they rated the emotions they perceived in both their responses and those of their peers. Emotional perceptions were assessed across four emotions (two negative: anger, anxiety; and two positive: happiness, hopefulness) using a 5-point Likert scale \cite{likert1932technique}. Additionally, we measured topic involvement with four items using an established scale \cite{chi13-involvement}.}
As outlined in Table \ref{table:prea}, the findings indicate that the topic of abortion resonated most strongly with our participants, eliciting the highest level of engagement. Consequently, we selected abortion as the central theme for our main study. Our rationale was that a topic eliciting such strong engagement would provoke more profound emotional reactions in our controlled laboratory setting. \rev{While we acknowledge that using a different topic may influence the specific emotional responses elicited, choosing abortion as the focal topic provided a robust context for investigating interaction effects due to its known ability to generate intense engagement. Moreover, abortion is not only a contentious issue in the US \cite{ALTSHULER2015226} but also globally \cite{reproductive19}, making it relevant and generalizable for studying social media interactions.} This, we anticipated, would create a more authentic scenario, enhancing the external validity of our data analysis.

\subsection{Emotional Effect}
\label{sec:effect}

\begin{wrapfigure}{r}{0.45\textwidth}
\centering
\frame{\includegraphics[width=0.45\columnwidth]{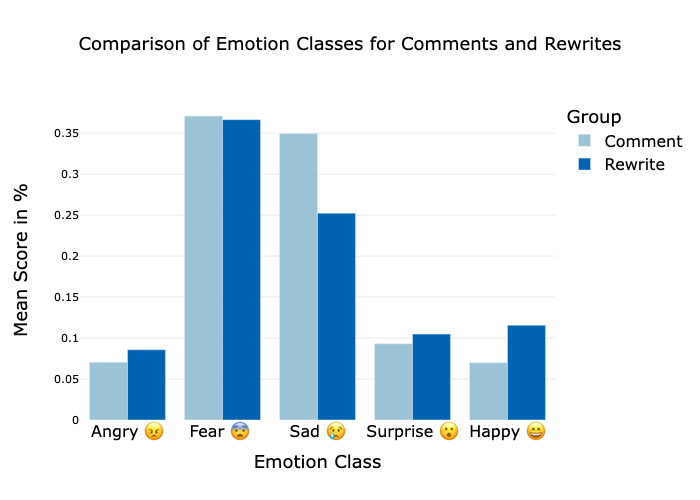}}
 \caption{Comparison of the mean value of emotion scores between original comments and rewritten texts detected by the \texttt{text2emotion} package.}
 \label{fig:preb_t2e}
\end{wrapfigure}
\subsubsection{Procedure} We recruited 67 participants (M (age) = 28.3, SD (age) = 11.5;  52.4\% male, 40.2\% female, 7.3\% other) from Prolific. Adhering to the result from the topic involvement experiment, we narrowed our focus to abortion as the sole topic of discussion. Following their posts, participants were prompted to evaluate the emotions in both their own and their peers' comments. This emotion rating task was intended as a stimulus to increase participants' emotional awareness and encourage more positive commenting behaviors on social media. Finally, we asked participants to revise their initial comments to a format they deemed appropriate for a social discussion. 

\subsubsection{Results}
Only messages exceeding 30 words and those participants that took more than 5 minutes to finish the survey were selected for the analysis. 
We used the \texttt{text2emotion} \footnote{\url{https://github.com/aman2656/text2emotion-library}} package and a fine-tuned Google T5 \footnote{\url{https://huggingface.co/mrm8488/t5-base-finetuned-emotion}} model to investigate the emotions. The \texttt{text2emotion} is a Python package that detects emotion-related words and categorizes them into five distinct emotions: Angry, Fear, Sad, Surprise, and Happy. The package outputs a dictionary for each message, with keys as emotion types and values as emotion scores in percentages, indicating the dominant emotion. \rev{Detailed description of the Google T5 model see Appendix \ref{appendix:t5}.}

As we intended, our analysis revealed reductions in scores of Fear and Sad, and an increase in Surprise and Happy (Figure \ref{fig:preb_t2e}). 
The results of a Shapiro-Wilk test \cite{shapiro1965analysis} showed that our data was not normally distributed. 
Therefore, we performed the non-parametric \ac{MWU} test between the emotion scores on the original and rewritten comments. There was a significant decrease in Sad according to the \ac{MWU} test (p = .028, U = 2726.5). Appendix \ref{appendix:t5} shows more details about the Google T5 model. \mrev{The figure shows that after rewriting, the messages classified as Angry, Fear, and Sad each decreased by about five, while those conveying Happy increased from just over 20 to 35.} Combining both results, we could conclude that participants tend to rephrase their comments with balanced and positive language after being prompted to rate their own and peer's emotions. This supports the idea that fostering emotional awareness leads to more positive, constructive dialogue, especially on sensitive topics like abortion.

%% file: tables/demographics.tex
\begin{table*}[!htbp]
\centering
\begin{tabular}{p{2em}p{2.5em}p{3em}ccccccc}
\toprule
\multirow{2}{*}{\textbf{Study}}
 & \multirow{2}{*}{\textbf{Group}}& \multirow{2}{*}{\textbf{N}} & \multicolumn{3}{c}{\textbf{Gender}} & \multicolumn{2}{c}{\textbf{Age}} & \multicolumn{2}{c}{\textbf{Highest Education}} \\
\cmidrule(lr){4-6} \cmidrule(lr){7-8} \cmidrule(lr){9-10}
& & & Male & Female & Other & Mean & Std. & < BSc & >= BSc \\
\midrule
Pre 1 & - & 41 & 46.3\% & 48.8\% & 4.9\% & 36.9 & 12.6 & 34.1\% & 65.9\% \\
\midrule
Pre 2 & - & 67 & 52.4\% & 40.2\% & 7.3\% & 28.5 & 11.5 & 44.5\% & 55.5\% \\
\midrule
\multirow{3}{*}{Main} & CG & 39 & 20.5\% & 76.9\% & 2.6\% & 28.3 & 11.3 & 51.3\% & 48.7\% \\
& TG1 & \colorbox{tg1}{\mbox{A1$\sim$A28}} & 28.6\% & 71.4\% & 0.0\% & 27.9 & 7.5 & 50.0\% & 50.0\% \\
& TG2 & \colorbox{tg2}{\mbox{B1$\sim$B36}} & 36.1\% & 61.1\% & 2.8\% & 26.1 & 6.1 & 47.2\% & 52.8\% \\
\bottomrule
\end{tabular}
\caption{Demographics data of participants in two pre-studies (Pre 1: topic involvement in Section \ref{sec:topic}), Pre 2: emotional effect in Section \ref{sec:effect}) and the main study in Section \ref{sec:main}. Participants are denoted as \colorbox{tg1}{A1} to \colorbox{tg1}{A28} in TG1 and denoted as \colorbox{tg2}{B1} to \colorbox{tg2}{B36} in TG2.}
\label{tab:demographics}
\end{table*}

%% file: tables/prea.tex
\label{sec:topic}
\begin{table}[h]
\centering
\begin{tabular}{lccc}
\toprule
\diagbox{\bfseries Measurement}{\bfseries Topic} & \textbf{Abortion} & \textbf{Television} & \textbf{Vegetarian} \\
\midrule
Self-Assessment: Negative Emotion & \textbf{2.6} & 2.0 & 1.9 \\
Self-Assessment: Positive Emotion & 1.6 & 1.7 & 2.0 \\
Peer-Assessment: Negative Emotion & \textbf{3.3} & 3.1 & 2.0 \\
Peer-Assessment: Positive Emotion & 1.5 & 1.5 & 2.2 \\
\midrule
Outcome-Relevant Involvement & \textbf{5.0} & 4.2 & 4.4 \\
Income-Relevant Involvement & \textbf{4.7} & 4.2 & 4.6 \\
\bottomrule
\end{tabular}
\caption{Comparison of the mean value for six measurements regarding three different topics (abortion, television, and vegetarian). \rev{The first four items on emotion assessments were rated on a 1 - 5 Likert Scale (1: not at all, 5: extremely). The last two items on topic involvement were rated on a 1 - 7 Likert Scale (1: low, 4: neutral, 7: high).}}
\label{table:prea}
\end{table}

%% file: sections/04_main_study.tex
\section{Main study}
\label{sec:main}
\begin{figure}[!ht]
\centering
 \includegraphics[width=1\columnwidth]{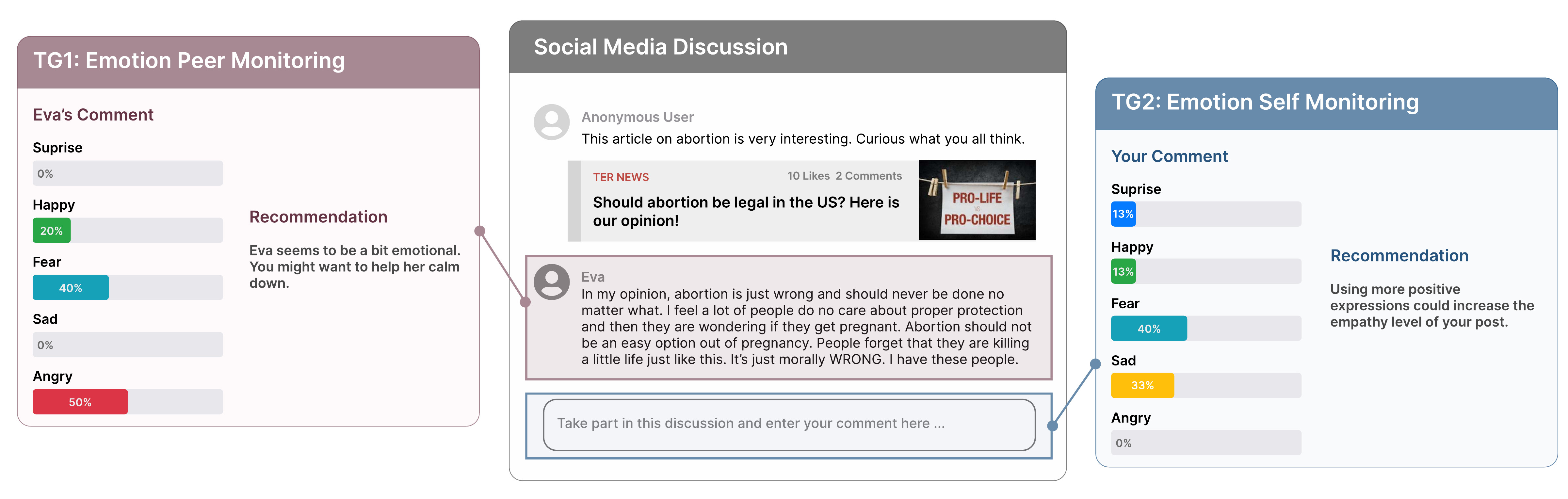}
 \caption{Overview of experimental setup in the main study: participants of the control group (CG) did not receive any analysis. Participants in treatment group 1 (TG1) received emotional analysis of a given peer's comment and participants in treatment group 2 (TG2) received emotional analysis of their own written comment. Both treatment groups receive personalized recommendations based on the emotional analysis. The interface of the emotion monitoring system is provided in Appendix \ref{appendix:interface}.}
 \label{fig:main-groups}
\end{figure}

The two pre-studies help us identify (1) topics that can be controversial and emotional, and (2) the potential of increased emotional awareness as a strategy to mitigate toxic communication.
Based on the results from the pre-studies, we conducted a controlled experiment with 103 participants posting comments after viewing a screenshot of a social media discussion on a controversial topic. 
Participants were randomly assigned to one of three conditions (as shown in Figure \ref{fig:main-groups}): the control group (CG), which did not receive any emotion monitoring; treatment group 1 (TG1), which received emotional analysis of a peer's comment; and treatment group 2 (TG2), which received emotional analysis of their own written comment.

Our goal was to gain empirical insights into how the two types of emotional monitoring influence users' communication patterns in social media discourse. 
Consequently, our analysis focused on comparing each treatment condition with the control condition (CG vs. TG1, CG vs. TG2) rather than comparing all three conditions against each other. 
Notably, these two mechanisms of emotional monitoring may coexist and complement each other when integrated into social media platforms.

\subsection{Participants and Task}
A total of \mrev{161} participants were recruited from the Prolific\footnote{\url{https://www.prolific.com/}} platform (M (age) = 27.4, SD (age) = 8.7; 28.2\%  male; 69.9\% female; 1.9\% other), as shown in Table \ref{tab:demographics}. 
Our objective was to obtain a representative subset of social media users, with a focus on younger individuals, while keeping the study open to participants of all ages.
Our study involves the following three phases:

\textit{Phase I.} We informed participants that they would engage in an online study regarding social media discussion and that their task was to write a comment. 
Then, we asked participants to fill out the \ac{PANAS} scale \cite{watson1988development}, openness to new experiences, and moral foundations. 
The \ac{PANAS} scale is a questionnaire assessing positive and negative affect, with scores measuring the level of agreement with each of the listed positive (active, attentive, determined, inspired) and negative (afraid, alert, hostile, nervous, upset) words on a 7 point Likert scale (1: totally disagree to 7: totally agree). This supported us to check people's initial emotional affect as a control mechanism. %

\textit{Phase II.} 
Mirroring the experimental paradigm of pre-studies, users were presented with a social media discussion centered on the topic of abortion. 
In this discussion, an initial comment is made by a user named \say{Eva}, \rev{who states that \say{In my opinion, abortion is just wrong and should never be done no matter what. I feel a lot of people do not care about proper protection and then they are wondering if they get pregnant. Abortion should not be an easy option out of pregnancy. People forget that they are killing a little life just like this. It's just morally WRONG. I have these people.}}
Then, participants are prompted to engage in the dialogue by responding to Eva’s perspective.
Upon drafting their response, they can press the button either \say{Emotion Peer Monitoring} (TG1) or \say{Emotion Self Monitoring} (TG2) to receive emotional feedback on the message they wrote. 
The subsequent dashboard provides an emotion analysis, showcasing a breakdown of five emotions in percentages on the left. 
For TG1, it demonstrates an emotional analysis of Eva's comment and TG2 would see an emotional analysis of their own comment. 
On the right, TG1 is encouraged to address Eva's message to make her less emotional and TG2 is offered personalized recommendations, guiding them in crafting more balanced and neutral comments. 
Appendix \ref{appendix:interface} displays the complete user interface of our developed tool. 

\textit{Phase III.} After posting comments, participants were asked to evaluate the peer's emotions according to the post and their own feelings about the previous discussion with several items. 
The answers were given on a 5-point Likert scale (1: not at all, 5: extremely) about five emotions (Angry, Fear, Sad, Surprise, and Happy). 
Next, we controlled for a battery of psychological mechanisms that could potentially \rev{interfere} with our results comprising of perceptions of topic involvement, communication efficiency, communication fairness, communication effectiveness, and elaboration. 
Finally, to receive fine-grained information on user experience with the emotion monitoring dashboard, we included three open-ended survey questions about their likes, improvement areas, and additional ideas in social media interactions.

\subsection{The Emotion Monitoring Dashboard}
We designed an emotion monitoring dashboard for presenting the peer and self monitoring feedback.
To inform the design of the emotion dashboard, we drew on an extensive review of existing emotion monitoring tools \cite{ELMIGID2022111206, EZZAOUIA2020102411, emoda17} and literature on emotion visualization for user interpretation \cite{torkildson2014analysis, kempter2014emotionwatch, 7042496}. 
In addition, we organized a design workshop, guided by the principles of research-through-design~\cite{chi07-interaction-design}, aiming at crafting a user interface that effectively supports emotion monitoring in social media interactions. 

\subsubsection{Design Workshop for the Emotional Dashboard}
We recruited 15 participants from diverse disciplines using mailing lists and leveraging internal university networks. There were seven females and eight males (M (age) = 28.7, SD (age) = 2.4), mainly HCI and UX researchers and practitioners (more demographic details see Appendix \ref{appendix:demographics}). 
The workshop spanned approximately 90 minutes and unfolded in three phases: \textit{(1) \textit{Identifying requirements}}: We established the dashboard's primary purpose, focusing on the need for effective emotion monitoring in social media discussions. \textit{(2) Detailing requirements}: We refined our initial requirements and deliberated on specific components of the user interface. \textit{(3) \textit{Implementing requirements}}: Our choice of bar graphs to represent sentiment data, with five distinct numerical values, was driven by clarity and immediate user comprehension. Additionally, we incorporated explanatory messages in personalized recommendations to boost user acceptance and understanding, promoting empathetic and reflective content engagement \cite{KAZIENKO202343}. 

\subsubsection{The Design of the Dashboard}
The outcome of the workshop was a user-centric emotion monitoring dashboard, designed to be intuitive and insightful for users engaging in social media discussions. 
The bar graph format for representing sentiment data was chosen for its clarity and immediate comprehension. 
The inclusion of explanatory messages was a strategic decision aimed at fostering a more empathetic and reflective interaction with the content. 
This comprehensive approach ensured that the dashboard not only met the identified user requirements but also aligned with the theoretical and practical aspects of emotion monitoring in digital communications.

\subsection{Data Analysis}
To examine the impact of emotion peer monitoring and emotion self monitoring, we conducted both quantitative and qualitative analyses.
Our measurements include the following.

\subsubsection{Data Quality Check}
To ensure high data quality, we only kept participants who passed the attention and manipulation checks. 
For the manipulation check, we asked participants to recall whether they received additional information when contributing to the discussion and what kind of information they received (e.g. on the emotionality of Eva’s comment or their own comment). 
Those who failed to choose the right answer were excluded from the final analysis (N = 21 for CG, N = 22 for TG1, N = 15 for TG2). 
Furthermore, we again only considered participants when their messages exceeded 30 words and when they took more than 5 minutes to finish the survey. \rev{This resulted in a final valid sample of 103 participants (N = 39 for CG, N = 28 for TG1, N = 36 for TG2), which we used to report our main findings}
Finally, we verified the \ac{PANAS} scores for the chosen users, ensuring they had comparable emotional levels at the beginning of the study and that there were no excessively emotional participants.

\subsubsection{Measured Emotions and Hate Speech in Users' Comments}
To analyze emotions in the written messages, we applied the \texttt{text2emotion} package and \ac{LIWC} \cite{pennebaker2001linguistic} with the newest dictionary\footnote{\url{https://www.liwc.app/}} to examine the emotional states in the written message. \rev{LIWC is a tool used for text analysis, particularly in psychological and linguistic research. It analyzes written text to identify emotional, cognitive, and structural components, categorizing words into various psychological and linguistic categories.} The results of measured emotions in the user comments are presented in Table \ref{tab:t2e_summary}, Figure \ref{fig:measured_tg1} and \ref{fig:measured_tg2}. Furthermore, we measured hate speech using the finetuned \texttt{Twitter-roBERTa-base} model \cite{acl20-tweeteval}\footnote{\url{https://huggingface.co/cardiffnlp/twitter-roberta-base-hate}}. \rev{This is a roBERTa-base model trained on around 58M tweets and finetuned for hate speech detection with the TweetEval benchmark. This model is specialized to detect hate speech against women and immigrants. Given a message, it classifies the content by determining the percentage of hate speech.} The result is presented in Figure \ref{fig:hate_all}.

\subsubsection{Linguistic Statistics in Users' Comments}
To examine the linguistic profile of happy and sad expressers, the written messages were analyzed using \ac{LIWC}. It analyzes messages on a word-by-word basis and compares words against a dictionary of words divided into different linguistic dimensions, including pronouns, affect terms, cognition terms, social and communicative processes. 
Its psychometric properties and external validity have been established in many studies, and have been used to examine the relationship between language and emotion. 
For the current study, only the variables relevant to our investigation were included in the analysis. 
The results of the linguistic profile are presented in Table \ref{tab:linguistic_tg1}, and \ref{tab:linguistic_tg2}.

\subsubsection{Self-Reported Emotions} Additionally, we analyzed user-reported emotions regarding both their peers and themselves, focusing on how users perceived their peers' emotions as well as their own emotional experiences.
The results are shown in Figure \ref{fig:peer_tg1}, \ref{fig:self_tg1}, \ref{fig:peer_tg2}, and \ref{fig:self_tg2}.

\subsubsection{Qualitative Analysis on the Perceptions of Emotional Monitoring} %
To probe into the perceptions of emotional monitoring, we conducted a qualitative analysis for the final open-ended questions. 
Based on qualitative analysis guidelines \cite{irr19}, we prioritized the discovery of common themes over strict inter-rater reliability, acknowledging that different coders might interpret meanings in varied ways \cite{irr19}. 
The first author and another researcher conducted reflexive thematic analysis \cite{braun2019reflecting} to code and develop themes. 
This process started with descriptive coding of user evaluations of emotion monitoring, followed by categorizing these codes into themes. 
Both researchers independently coded the entire dataset and then cross-checked each other's work. 
A third researcher validated and finalized these themes.

%% file: sections/05_result.tex
\section{Results}
We organized the findings for the two research questions based on four main themes: (1) participants' perceptions of the monitoring dashboard, (2) measured emotions and hate speech in user comments, (3) the reported emotions of their peers and their own, and (4) the difference in the linguistic usage of treatment groups compared to the control group.
\input{tables/t2e_summary_new}

\begin{figure}[!ht]
\centering
\frame{\adjustbox{padding=10pt}{\includegraphics[width=.8\columnwidth]{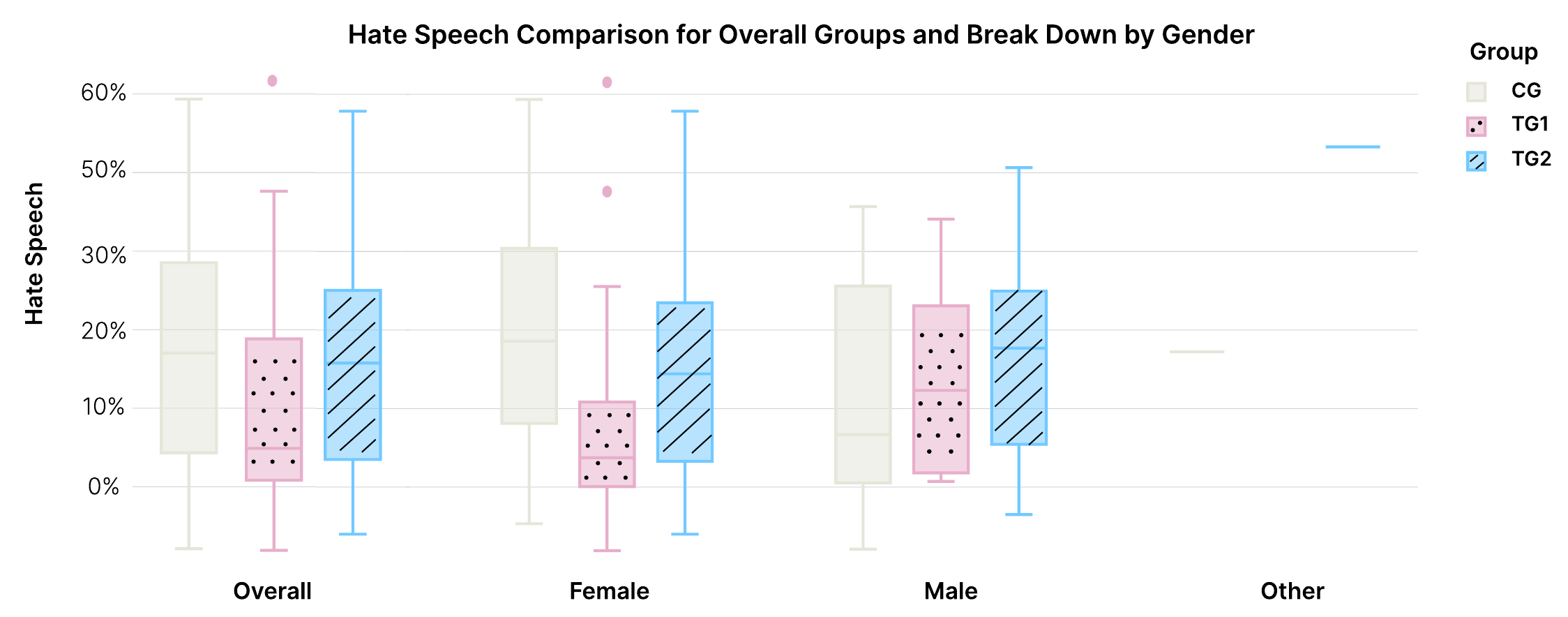}}}
 \caption{Hate speech comparison for all three groups (CG, TG1, and TG2), using \texttt{Twitter-roBERTa-base}.}
 \label{fig:hate_all}
\end{figure}

\subsection{RQ1: Effects of Emotion Peer Monitoring}
\subsubsection{Participants thought the analysis was accurate and tried to adjust their wording to empathize with others.}

In TG1, we presented participants with the emotion analysis of their peer Eva's message. \rev{71.4\% of the participants expressed their appreciation of the breakdown of emotions and real-time analysis} (e.g. \colorbox{tg1}{A17}: \say{\textit{I liked the breakdown of the emotions which demonstrates the complexity in people's responses}}, \colorbox{tg1}{A9}: \say{\textit{The system is very intriguing. I’ve never seen anything like that before. I think it will serve people well in learning about themselves and their reactions to posts on social media.}}). \rev{Three participants (\colorbox{tg1}{A3}, \colorbox{tg1}{A8}, \colorbox{tg1}{A27}) explicitly expressed that they thought the analysis was accurate and confirmed their feelings about the peer's comment, for example, \colorbox{tg1}{A8} wrote: \say{\textit{I felt that it accurately portrayed the emotions of the other commenter. I firmly disagree with their viewpoint. The feedback system validated my suspicion that the commenter was scared and angry.}}.} Only two participants wrote that the system was inaccurate (\colorbox{tg1}{A31}, \colorbox{tg1}{A55}). Moreover, the system's ability to facilitate empathetic communication was highlighted by several participants. For instance, \colorbox{tg1}{A18} mentioned \say{\textit{Because I knew how people felt about Eva's comment, I tried to put more thought into my comment and tried to be mindful of the views of both sides of the discussion.}}. Interestingly, participants also noted that the emotion analysis contributed to their self-reflection. \colorbox{tg1}{A19} stated \say{\textit{It reminded me to take a step back and frame my response in a way that *maybe* Eva could actually hear me. It felt like I was given a little insight into how to build a better bridge, rather than burn the whole thing down.}} These insights underscore the system's potential to foster more constructive and empathetic online interactions by making users aware of the emotional undertones of communications.

\subsubsection{Heightened negative emotions, reduced positive emotion and reduced hate speech were detected in the message from female participants, while the reversed trend was detected in the message from male participants.}
\label{sec:6.1.2}
In their final comments, as shown in Table \ref{tab:t2e_summary}, the message written by TG1 exhibited a higher inclination towards all negative emotions (Angry, Fear, and Sad), coupled with a reduction in neutral (Surprise) and positive (Happy) emotions. However, none of these differences were statistically significant. Since over two-thirds of the participants in the main study were female, the overall trends were predominantly influenced by female responses. Analyzing the data by gender, as illustrated in Figure \ref{fig:measured_tg1}, reveals divergent trends among different gender groups. Due to the small number of participants identifying as \say{Other}, we only consider binary gender in this analysis.

The figure indicates a general increase in negative emotions (Angry, Fear, Sad) and a decrease in Happy among female participants, whereas male participants exhibited opposite trends: a reduction in Angry and Fear, and an increase in Happy. Notably, we observed a significant reduction in Happy for females (Two-sample T-test, \textit{p} = .0264, T = 2.29) and a significant increase in Sad (\ac{MWU} test, \textit{p} = .0193, U = 182). Despite the intensified negative emotions among female participants, their comments contained less hate speech, whereas hate speech increased among male participants, as shown in Figure \ref{fig:hate_all}.

It is also worth noting that compared to the control group, TG1 has a reduced hate speech, and this is statistical significance according to the \ac{MWU} test (\textit{p} = .0257, U = 722). Examining the reduction of hate speech from a gender perspective gives us another insight. From Figure \ref{fig:hate_all} we can see that females reduced hate speech TG1 while males slightly increased hate speech. And the reduction of hate speech of females in TG1 is significant (\textit{p} = .00647, U = 438).

\input{tables/fig_measured_tg1}
\input{tables/fig_peer_tg1}
\input{tables/fig_self_tg1}

\subsubsection{Participants used more pronouns and fewer articles.}
To further understand the difference, we \rev{tried} to examine their use of languages using LIWC. From linguistic dimensions, we found that compared to CG, TG1 participants used pronouns more frequently and articles less frequently. When we \rev{examinedd} further, we discovered that this statistical significance actually came from female participants as shown in Table \ref{tab:linguistic_tg1}. Even though there were similar trends in male participants, the effects were not as significant.
\rev{
The use of personal pronouns is an important aspect of language that reveals information about people's mental states \cite{LYONS2018207}. Pronouns indicate whether an individual's focus is on the self (first-person singular pronouns such as “I”, “me”, “my”), on others (second-person pronouns such as “s/he”, “you”, “they”, “others”). 
Pronouns like ``I'' and ``you'' are often used to establish connections between interlocutors, with their increased usage reflecting a more engaged and personalized discourse.
Specifically, the frequent use of first-person singular pronouns signals a higher degree of self-reference, often associated with greater emphasis on personal experiences, emotions, or perspectives.
Research suggests that this inward focus, as evidenced by the elevated use of first-person singular pronouns, can be indicative of mental distress \cite{EDWARDS201763, zimmermann2017first}. This aligns with the heightened negative emotions observed in the messages from female participants, as discussed in Section \ref{sec:6.1.2}.
}
Alternatively, this change of wording may reflect a shift towards more direct and concise language. Articles like ``the'' and ``a'' are often used to specify or generalize nouns, and their reduced frequency could indicate a preference for brevity or a more streamlined communication style. This might suggest that individuals in TG1 are more focused on conveying their emotions or ideas directly, without excessive elaboration. 
The observed changes in pronoun and article usage could be attributed, at least in part, to the emotional states experienced by individuals in the treatment group. Negative emotions, such as Sad or Angry, might lead to a greater emphasis on personal experiences and perspectives, as well as a more direct expression of thoughts or feelings. 
\input{tables/linguistic_tg1}

\subsubsection{Participants did not view the peer's emotions differently compared to CG and they viewed their own emotions less negatively.}
Even though TG1 participants were presented additionally with the emotion analysis of Eva's comment compared to CG, there was not much difference in how they viewed the emotions in Eva's comment, no matter whether overall or by gender as shown in Figure \ref{fig:peer_tg1}. However, in their self-reported emotions, both female and male participants reported they experienced less negative emotions, as shown in Figure \ref{fig:self_tg1}.

\subsubsection{Summary of Findings for RQ1.} The findings suggest that the exposure of participants in TG1 to emotion analysis of peer messages influenced their perception and expression of emotions. This influence is reflected not only in the evaluation of peer comments but also in the linguistic features of their communication. The reduced hate speech and qualitative feedback indicating the accuracy of the analysis further underscore the role of emotion monitoring in shaping language use and interpersonal communication.

\subsection{RQ2: Effects of Emotion Self Monitoring}

\subsubsection{Participants thought the analysis was inaccurate and reported the discrepancy between emotional feedback and their true feelings.}
In TG2, we presented participants with the emotion analysis of their own written messages. In contrast to the positive feedback from TG1, there are more polarized opinions on the emotion self monitoring system in TG2. While \rev{over half of the participants (58.3\%)} in TG2 expressed appreciation for the system and the concept of emotion monitoring, concerns were raised about its accuracy and the alignment of the analysis with their personal feelings. \rev{Seven participants complained that they did not think the system was accurate (\colorbox{tg2}{B4}, \colorbox{tg2}{B8}, \colorbox{tg2}{B12}, \colorbox{tg2}{B15}, \colorbox{tg2}{B19}, \colorbox{tg2}{B29}, \colorbox{tg2}{B33}). Additionally, four participants (\colorbox{tg2}{B15}, \colorbox{tg2}{B21}, \colorbox{tg2}{B26}, \colorbox{tg2}{B34}) reported discrepancies between the emotions portrayed in the dashboard and their actual emotional experiences (e.g. \colorbox{tg2}{B34}: \say{I'm not really sure how accurate it was. Yes, sadness was a big part of my response, but I felt anger was more dominant in my response.}), potentially leading them to disregard the dashboard's feedback.
Furthermore, two participants (\colorbox{tg2}{B1}, \colorbox{tg2}{B29}) found it challenging to reduce the negative aspects of their language. \colorbox{tg2}{B1} noted \say{\textit{It was difficult to tone down my emotional statuses. I did not think I was being negative, but that is what the emotion dashboard determined.}}}

This discrepancy may be due to the following factors. Firstly, the algorithm's inherent limitations in capturing nuanced contextual cues led to reduced accuracy and made people disregard the feedback. The \texttt{text2emotion} package works by detecting emotion-related words and categorizing them into five predefined emotions. When participants discuss negative topics, the algorithm reports negative emotions even if the participants do not intend to express such emotions. For instance, \colorbox{tg2}{B17} stated \say{\textit{In my original comment, I talked about rape situations where abortion is necessary. When I took out the word rape, my fear score went down. Yes, rape is bad, but I'm not actively afraid, I'm just talking about the action as a simple fact.}}. Therefore, there might be a disconnect between the emotions users intend to convey and those that are effectively communicated through their messages. This misalignment highlights the complex interplay between users' emotional intentions and the expressive nuances of their language. 

\subsubsection{Heightened negative emotions and a reduced positive emotion were detected in the message from both females and males, but only female participants reduced hate speech.}

Figures \ref{fig:measured_tg2} and Table \ref{tab:t2e_summary} present that more negative emotions and less positive emotions were detected in messages from both male and female participants. However, in Figure \ref{fig:hate_all}, we can see that only female participants had a reduced hate speech.

\begin{center}
\input{tables/fig_measured_tg2}
\end{center}

\begin{center}
\input{tables/fig_peer_tg2}
\end{center}

\begin{center}
\input{tables/fig_self_tg2}
\end{center}

\subsubsection{Participants used more auxiliary verbs and fewer conjunctions.}
Regarding language usage, TG2, similar to TG1, demonstrated a higher frequency of pronoun usage compared to the control group, although this difference was not statistically significant (as shown in Table \ref{tab:linguistic_tg2}). Conversely, TG2 displayed lower usage of articles, but again, this difference did not reach statistical significance. 
However, according to the \ac{MWU} test, TG2 did exhibit a significant increase in the usage of auxiliary verbs (p = .0161, U = 475) and a decrease in conjunctions (p = .0175, U = 927) compared to the control group. These results suggest a potential shift towards more direct and action-oriented language within TG2.

The increase in auxiliary verb usage in TG2 may indicate a tendency towards expressing actions, states, or conditions more explicitly. Auxiliary verbs often serve to clarify tense, aspect, modality, and voice, which can make language more precise and direct. For instance, participants in TG2 might be more focused on describing their current emotional states or anticipated actions, reflecting the impact of the emotion self monitoring system on their language patterns. \rev{Alternatively, one hypothesis could be that the emotion self monitoring system encouraged participants to narrate their experiences in a step-by-step or instructional manner, which often relies on auxiliary verbs to describe processes and sequences.}
The decrease in conjunction usage in TG2 suggests a reduction in the complexity of sentence structures. Conjunctions are used to link clauses, phrases, and words, creating more complex and compound sentences. A reduction in conjunction usage may indicate a shift towards simpler, more straightforward communication. This could be a result of participants attempting to convey their emotions more clearly and directly, possibly in response to the feedback from the emotion self monitoring system. \rev{Alternatively, this simplification could result from cognitive load: participants might subconsciously reduce sentence complexity when focusing on emotional regulation tasks, prioritizing clarity over nuance.}

\subsubsection{Participants view the peer's emotions more negatively compared to CG and they viewed their own emotions differently for different genders.}

From Figure \ref{fig:peer_tg2} we can see that compared to CG, TG2 participants evaluate their peer's message had more negative emotions for both males and females.
For self-reported emotions, there was a gender difference. Female participants reported experiencing more negative emotions while male participants reported experiencing less negative emotions as shown in Figure \ref{fig:self_tg2}. 
This trend is reflected in their messages, where more negative emotions were evident, and fewer positive emotions were expressed, as detailed in Table \ref{tab:t2e_summary}. However, statistical analysis revealed that these emotional differences did not reach statistical significance when compared between the groups. 

\input{tables/linguistic_tg2}
\subsubsection{Summary of Findings for RQ2.} 
The findings suggest that the exposure of participants in TG2 to emotion analysis of their own messages influenced their perception and expression of emotions. We observed a reversed trend in how female and male participants evaluate their own emotions. Furthermore, we found the reduction in hate speech and significant changes in linguistic features both came from female participants.

\subsection{\rev{Transparency and censorship}}
\rev{In addition to the findings from isolated factors, several common concerns emerged across both participant groups. Despite efforts to enhance user comprehension and acceptance through personalized recommendations and explanatory messages, many participants expressed a lack of transparency in explaining how emotional scores are calculated. Participants from both groups emphasized the need for clearer communication regarding the methodology. 
Over one-third participants in TG1  (\colorbox{tg1}{A1}, \colorbox{tg1}{A2}, \colorbox{tg1}{A4}, \colorbox{tg1}{A7}, \colorbox{tg1}{A10}, \colorbox{tg1}{A12}, \colorbox{tg1}{A15}, \colorbox{tg1}{A16}, \colorbox{tg1}{A17}, \colorbox{tg1}{A25}) and four participants in TG2 (\colorbox{tg2}{B7}, \colorbox{tg2}{B12}, \colorbox{tg2}{B17}, \colorbox{tg2}{B30}) expressed that they struggled to interpret the results and requested insight into the algorithm's workings.
For instance, from TG1 \colorbox{tg1}{A12} mentioned \say{\textit{It could give more information on how it worked, I was still confused after I looked at the explanation.}} and \colorbox{tg1}{A16} suggested \say{\textit{Make the methodology more transparent. Even just a blurb explaining how it works.}} Similarly, \colorbox{tg2}{B7} in TG2 noted \say{\textit{Highlight or underline which words or sections of a comment are triggering certain positive or negative responses from the feedback tool.}} This lack of transparency may not only impede users' ability to act upon emotional feedback but also contribute to skepticism and disengagement with the system. As \colorbox{tg1}{A10} noted \say{\textit{I did not understand what I was supposed to do. It was confusing and made me feel frustrated.}}
}

\rev{
Interestingly, participants in TG2 who received emotion self monitoring features reported feeling censored—a perception not observed among TG1 participants. For example, TG2 participant \colorbox{tg2}{B13} remarked \say{\textit{Rather than feeling like I'm making sure my comments are civil, it really just feels like I'm being censored.}} Similarly, \colorbox{tg2}{B32} expressed skepticism about the system’s utility, stating \say{\textit{It's extremely hectoring and judgmental. I don't need my sentiment to be analyzed since obviously I know how I feel (and better than an algorithm).}}}

%% file: tables/t2e_summary_new.tex
\begin{table}[!h]
\small
\centering
\begin{tabular}
{p{2.8em}p{3em}p{3em}p{3em}p{3em}p{3em}p{3em}p{3em}p{3.2em}p{3.2em}}
    \toprule
    \multirow{ 2}{*}{\textbf{Group}}
    & {\scriptsize Angry} & {\scriptsize Fear} & {\scriptsize Sad} 
    & {\scriptsize Surprise}
    & {\scriptsize Happy}
    & \multicolumn{2}{c}{\textbf{Emotion}} & \multicolumn{2}{c}{\textbf{Tone}} \\
    & {\scriptsize Angry} & {\scriptsize Fear}& {\scriptsize Sad} & {\scriptsize Surprise} & {\scriptsize Happy} & {\scriptsize Negative} & {\scriptsize Positive} & {\scriptsize Negative} & {\scriptsize Positive}\\
    \midrule
    CG  & 6.6\% & 8.0\% & 14.1\% & 25.8\% & 45.7\% & 0.18\% & 0.64\% & 4.92\% & 1.81\%\\
    \midrule
    TG1 & 7.4\% & 8.6\% & 22.1\% & 21.4\% & 40.4\% & 0.21\% & 0.69\% & 4.81\% & 1.38\% \\
    \midrule
    TG2 & 10.9\% & 7.9\% & 20.2\% & 21.4\% & 39.3\% & 0.54\% & 0.43\% & 4.96\% & 2.30\%\\
    \bottomrule
\end{tabular}
\caption{Statistics summary of the percentages of five measured emotions using the \texttt{text2emotion} package, and the overall tone and emotion measurement using LIWC-22. The numbers in the last two columns represent the percentage of words in the comments that fall into the corresponding category. The positive and negative emotions are restricted to words that strongly imply corresponding emotions. The positive and negative tone dictionaries include words related to positive and negative emotions (e.g., happy, sad, angry) and also words related to those emotions (e.g., birthday, beautiful, kill, funeral) \cite{monzani2021emotional}.}
\label{tab:t2e_summary}
\end{table}

%% file: tables/fig_measured_tg1.tex
\begin{figure}
    \centering
    \includegraphics[width=\linewidth]{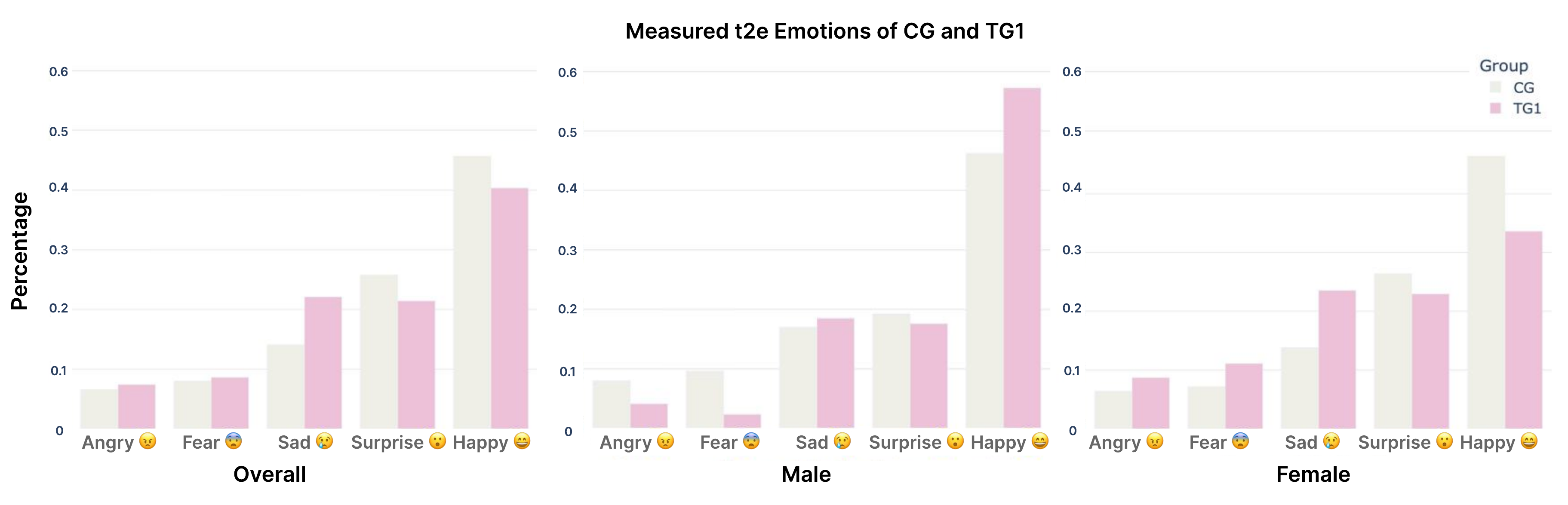}
    \caption{Comparison of measured emotions using the \textit{text2emotion} package between CG and TG1.}
    \label{fig:measured_tg1}
\end{figure}

%% file: tables/fig_peer_tg1.tex
\begin{figure}
    \centering
    \includegraphics[width=\linewidth]{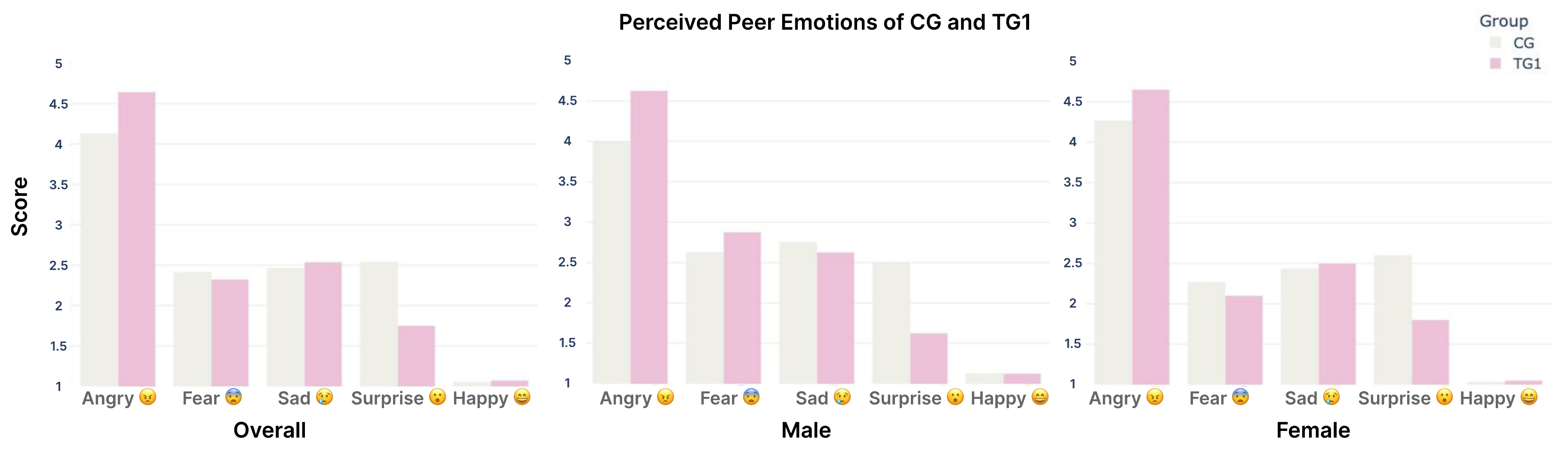}
    \caption{Comparison of perceived peer emotions between CG and TG1. Participants were asked to rate to what extent do they feel certain emotions from a 1-5 Likert scale.}
    \label{fig:peer_tg1}
\end{figure}

%% file: tables/fig_self_tg1.tex
\begin{figure}
    \centering
    \includegraphics[width=\linewidth]{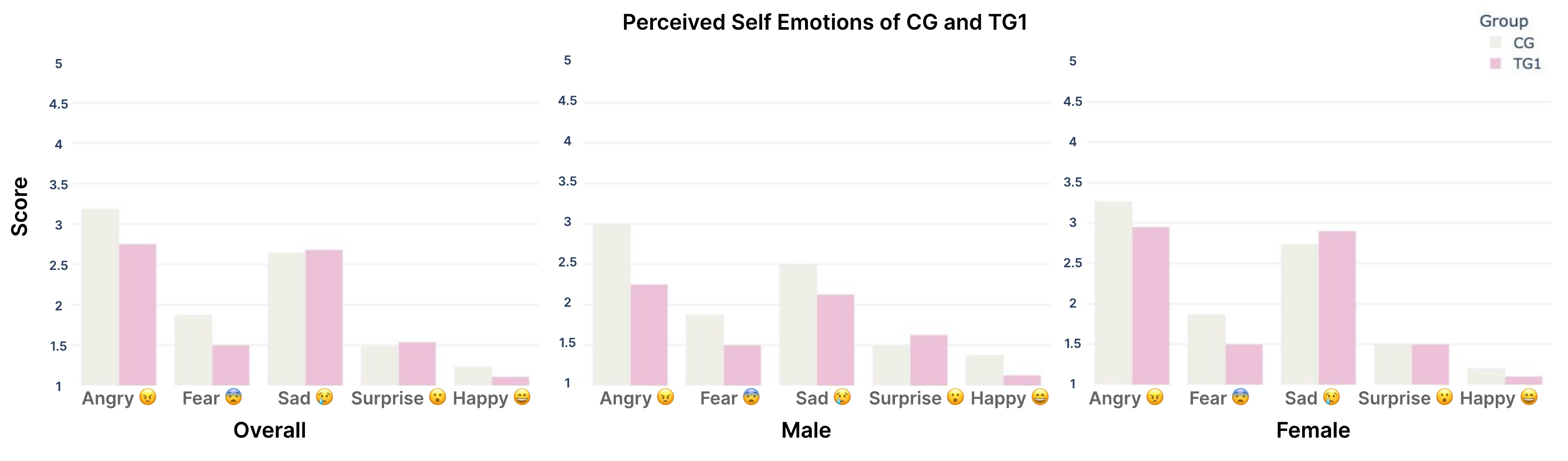}
    \caption{Comparison of perceived self emotions between CG and TG1. Participants were asked to rate to what extent do they feel certain emotions from a 1-5 Likert scale.}
    \label{fig:self_tg1}
\end{figure}

%% file: tables/linguistic_tg1.tex
\begin{table}[H]
\centering
\begin{tabular}{p{2cm}rrrrrrrr}
\toprule
\textbf{Group} & \textbf{pronoun} & \textbf{ppron} & \textbf{ipron} & \textbf{i} & \textbf{you} & \textbf{article} & \textbf{auxverb} & \textbf{conj} \\ \midrule
CG    & 12.43 & 6.56 & 5.88 & 1.81 & 1.01 & 7.51 & 12.05 & 7.77 \\
TG1   & 15.97 & 8.17 & 7.79 & 2.31 & 2.13 & 5.45 & 12.88 & 7.38 \\
\textit{p-value}  & \textbf{.005**} & \textbf{.023*} & \textbf{.023*} & \textbf{.038*} & \textbf{.018*} & \textbf{.011*} & .407 & .516 \\ \midrule
CG (female)  & 11.77 & 6.59 & 5.18 & 1.73 & 1.11 & 7.31 & 11.67 & 7.73 \\
TG1 (female) & 15.91 & 8.34 & 7.57 & 2.44 & 2.55 & 5.45 & 12.81 & 7.82 \\ 
\textit{p-value}  & \textbf{.003**} & .082 & \textbf{0.02**} & \textbf{.023} & \textbf{.036*} & \textbf{.042*} & .335 & .889 \\ \midrule
CG (male)   & 14.37 & 6.13 & 8.24 & 2.31 & 0.21 & 8.99 & 13.28 & 7.96 \\
TG1 (male)  & 16.12 & 7.75 & 8.36 & 1.98 & 1.08 & 5.47 & 13.06 & 6.29 \\
\textit{p-value}  & .557 & .375 & .952 & .688 & .054 & .087 & .920 & .200 \\ \bottomrule
\end{tabular}
\caption{Comparison of linguistic statistics between CG and TG1. We used the Student's t-test for normally distributed groups, Welch's t-test for normally distributed but not equal variance groups, and used the MWU test for not normally distributed groups (*p < 0.05, **p < 0.01, ***p < 0.001).}
\label{tab:linguistic_tg1}
\end{table}

%% file: tables/fig_measured_tg2.tex
\includegraphics[width=\linewidth]{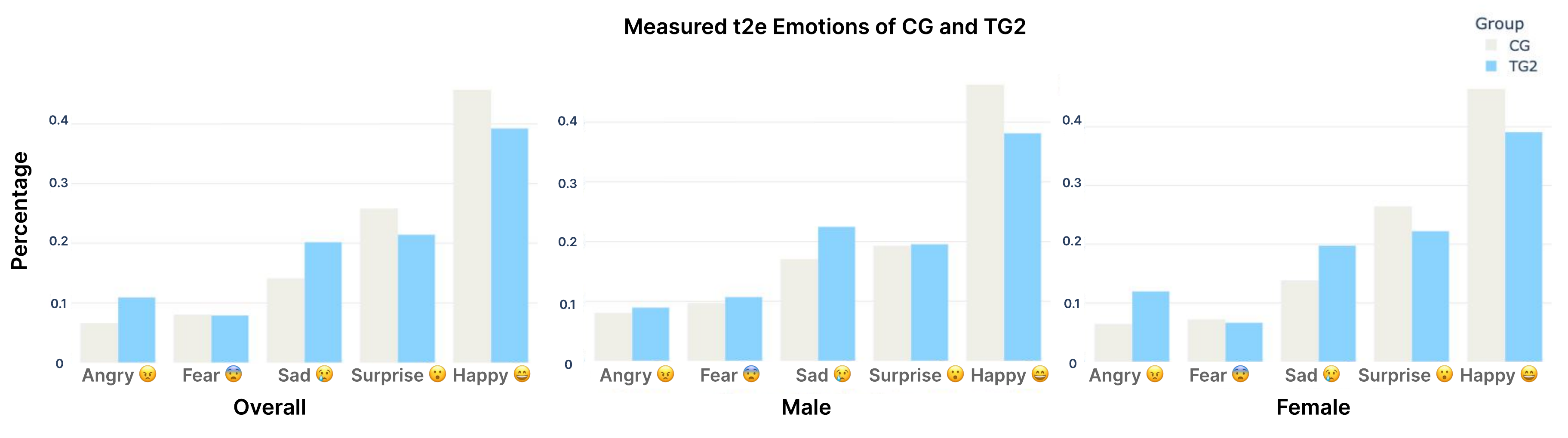}
\captionof{figure}{Comparison of measured emotions using the \textit{text2emotion} package between CG and TG2.}
\label{fig:measured_tg2}

%% file: tables/fig_peer_tg2.tex
\includegraphics[width=\linewidth]{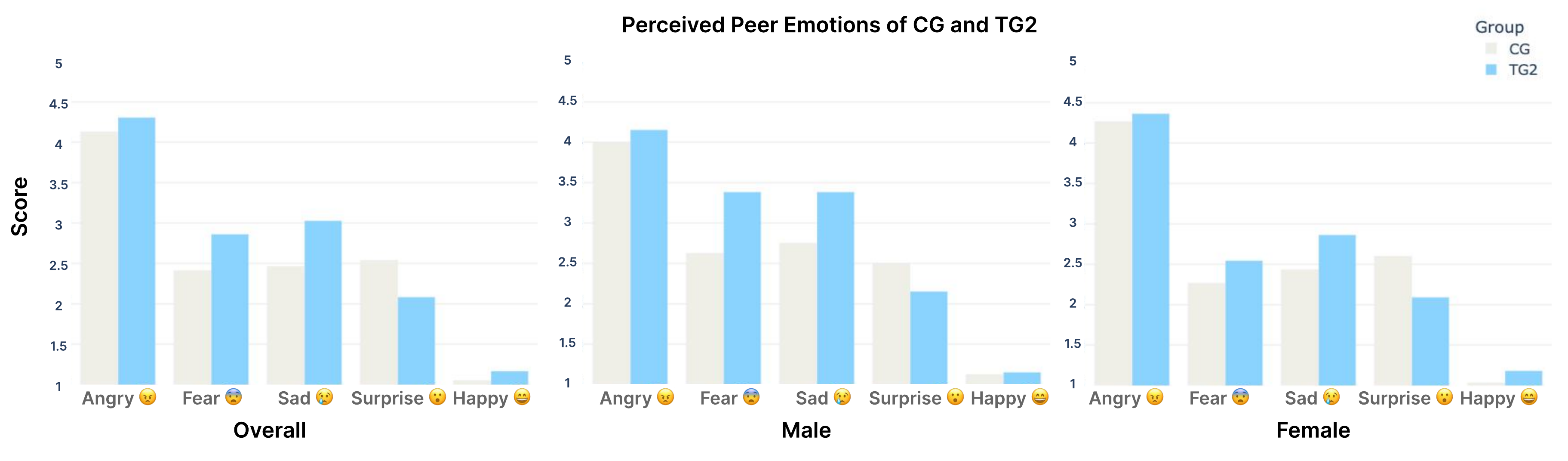}
\captionof{figure}{Comparison of perceived peer emotions between CG and TG2. Participants were asked to rate to what extent do they feel certain emotions from a 1-5 Likert scale.}
\label{fig:peer_tg2}

%% file: tables/fig_self_tg2.tex
\includegraphics[width=\linewidth]{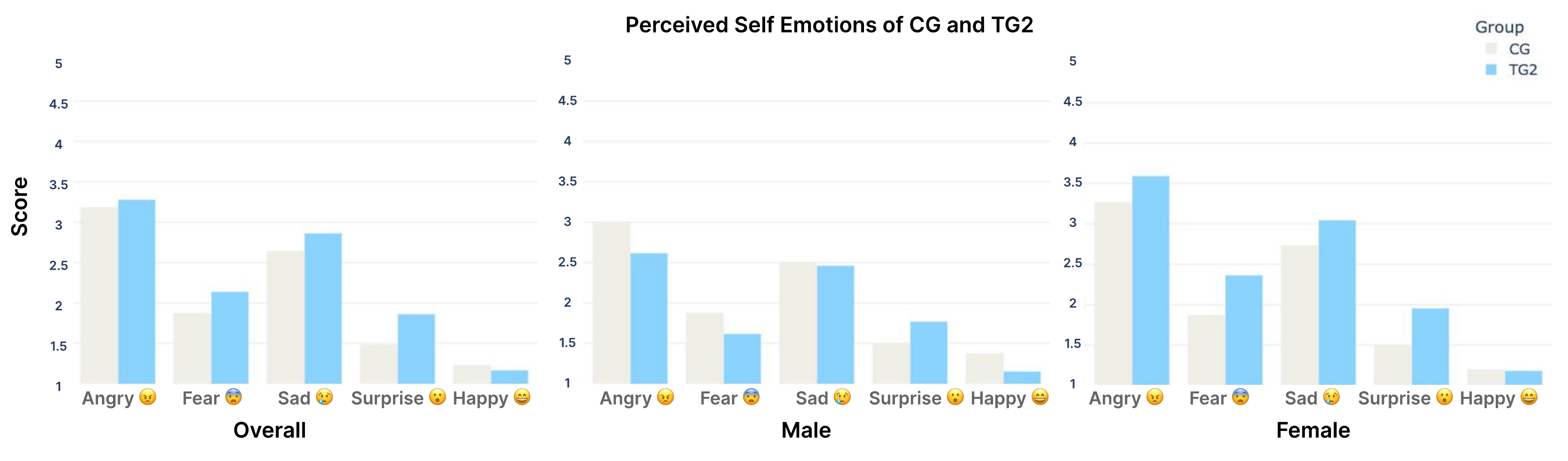}
\captionof{figure}{Comparison of perceived self emotions between CG and TG2. Participants were asked to rate to what extent do they feel certain emotions from a 1-5 Likert scale.}
\label{fig:self_tg2}

%% file: tables/linguistic_tg2.tex
\begin{table}[H]
\centering
\begin{tabular}{p{2.5cm}rrrrrrrr}
\toprule
\textbf{Group} & \textbf{pronoun} & \textbf{ppron} & \textbf{ipron} & \textbf{i} & \textbf{you} & \textbf{article} & \textbf{auxverb} & \textbf{conj} \\ \midrule
CG    & 12.43 & 6.56 & 5.88 & 1.81 & 1.01 & 7.51 & 12.05 & 7.77 \\
TG2  & 14.44 & 7.36 & 7.08 & 2.24 & 1.11 & 7.18 & 13.70 & 6.63 \\
\textit{p-value}  & .146 & .754 & .151 & .310 & .961 & .669 & \textbf{.016*} & \textbf{.035*} \\ \midrule
CG (female)  & 11.77 & 6.59 & 5.18 & 1.73 & 1.11 & 7.31 & 11.67 & 7.73 \\
TG2 (female) & 13.84 & 7.04 & 6.80 & 2.02 & 1.03 & 7.38 & 13.46 & 6.21 \\ 
\textit{p-value}  & .173 & .712 & .364 & .426 & .727 & .941 & \textbf{.038*} & \textbf{.013*} \\ \midrule
CG (male)   & 14.37 & 6.13 & 8.24 & 2.31 & 0.21 & 8.99 & 13.28 & 7.96 \\
TG2 (male)  & 15.99 & 8.03 & 7.96 & 2.79 & 1.33 & 7.11 & 14.03 & 6.85 \\
\textit{p-value}  & .622 & .375 & .952 & .689 & .054 & .09 & .385 & .358 \\ \bottomrule
\end{tabular}
\caption{Comparison of linguistic statistics between CG and TG2. We used the Student's t-test for normally distributed groups, Welch's t-test for normally distributed but not equal variance groups, and used the MWU test for not normally distributed groups (*p < 0.05, **p < 0.01, ***p < 0.001).}
\label{tab:linguistic_tg2}
\end{table}

%% file: sections/06_discussion.tex
\section{Discussion}
Through a series of studies, we investigated the effects of two types of emotion monitoring on emotional experience and linguistic behavior in online communication of a sensitive topic. Similar to \cite{cscw22-thread}, our findings indicate that dynamic, personalized interventions may substantially impact user's commenting behavior. Specifically, our findings show that emotional analytics may enhance emotional awareness in social media discussions. 

\subsection{Theoretical Implications}
Table \ref{tab:discussion} summarizes the findings by topic through two research questions. The findings indicate that exposure to emotion analysis of peer messages significantly impacts users' perception and expression of emotions. Participants of TG1 exhibited a heightened sensitivity towards negative emotions in peer comments. The qualitative feedback further corroborates these findings, emphasizing the role of emotion monitoring in fostering empathy and guiding users toward adjusting the emotional tone of their messages. This highlights the potential of emotion peer-monitoring to not only influence individual behavior but also facilitate more empathetic and emotionally attuned interactions within online discussions.

\begin{center}
\input{tables/discussion}
\end{center}
Interestingly, female participants in both monitoring groups expressed more negative emotions in their comments. We controlled for participants’ emotional states using the PANAS scores before they began the conversation. This ensured that all participants had comparable emotional baselines at the start, with no individuals exhibiting excessively high emotional levels. \mrev{While the heightened negative emotions among female participants could, in part, be attributed to the topic -—abortion. It is important to note that existing research suggests mixed findings regarding gender-based attitudes toward abortion.} Specifically, while abortion is a cornerstone issue in the women’s rights movement, attitudes toward it are not consistently aligned with gender \cite{osborne2022abortion, abortion15}. Besides different genders could have different attitudes towards abortion.

\mrev{Another explanation for the heightened negative emotions could lie in the differences in how males and females process emotional stimuli. Research indicates that females may engage different neurocircuitry than males during perceptual emotion processing. In some cases, this can lead to more accurate or faster emotional assessments, while in others, it may result in heightened reactivity, potentially increasing vulnerability to affective disorders \cite{journal98-sex}. }

This result also highlights a gendered dimension in emotional sensitivity and expression that warrants further exploration. The heightened levels of negative emotions, particularly Angry and Sad, underscore the potential impact of emotion monitoring interventions on emotional expression. These results align with previous research indicating that heightened self-awareness of emotions can sometimes lead to increased negative affect, especially when individuals become overly focused on monitoring and regulating their emotional experiences \citep{silvia2002self}. As heightened awareness reduces the inclination to suppress genuine feelings, leading to more pronounced negative emotions in responses \cite{authentic14}. Despite the increase in negative emotional expression, we also observed a decrease in the use of hate speech among these participants, indicating a complex interaction between emotional awareness and communicative behavior.

The linguistic analysis of the texts revealed that several verbal dimensions differed significantly across the two monitoring conditions. 
Both treatment groups (TG1, TG2) have an increase in the use of pronouns and auxiliary verbs, suggesting that emotion monitoring systems might encourage users to articulate their emotions and actions more clearly. This could be beneficial for improving emotional self-awareness and communication.

Additionally, the accuracy of the system's analysis was a key factor in its acceptance and effectiveness. While the majority of participants found the analysis accurate, those who did not raise concerns about the potential misinterpretation of emotions (\textcolor{tg1}{A31}, \textcolor{tg1}{A55}). 
The discrepancy between users' self-reported emotional states and the emotional content of their messages highlights the complexity of emotional expression in online communication. Users' attempts to adjust their messages to diminish perceived negativity suggest a conscious effort to align their online persona with their desired emotional presentation. However, the challenges they encountered in accurately conveying their emotional states suggest limitations in current emotion detection algorithms and the need for greater transparency in how emotional scores are computed.
This feedback also highlights the importance of continually refining the emotion analysis algorithms to ensure reliability and to address the nuances in human emotions effectively.

\subsection{Practical Implications}
This study holds practical implications for designers and researchers. First, it underscores the significant influence of platform design on user behavior \cite{KAZIENKO202343}. Specifically, designers should carefully identify features that support users in empathizing with their peers on these platforms \cite{soral2022role, cscw19-accountability} and build the self-reflection capacities of users \cite{royen2022think, talk20}. Second, as algorithmic outputs grow more complex, users require guidance on interpreting and acting on this feedback while the design of explainable systems \cite{explain18}. Trust in these algorithms hinges on transparent explanations of their workings, making such clarity essential for their acceptance \cite{cscw21-recast, helpai23}. As \citet{chi20-human} highlight, users often have limited understanding of how their emotions are detected and interpreted, which can lead to misinformed interactions with these tools. Designers must prioritize clear and accessible explanations of how these systems work, including the data sources, algorithms, and potential biases involved. Enhancing transparency in algorithmic interface \cite{cscw19-transparency, transparency16, cscw21-recast} could bridge this trust gap, helping users understand and accept these tools more readily. \mrev{Finally, these findings suggest that future investigations into emotion regulation must take a holistic, context-sensitive approach. Ultimately, by embedding these contextual considerations into both design and research, future developments in emotion regulation technologies can better support users’ well-being.}

In conclusion, the preliminary findings suggest that the emotion analysis system is not only well-received but also has a positive impact on users' online interactions. By providing insights into the emotional content of messages, the system encourages empathy, self-reflection, and more mindful communication. These results pave the way for further investigations into the long-term benefits of such systems in promoting healthier and more productive online discourse. Future work will involve larger-scale studies to validate these findings and explore additional features that could enhance the system's utility and user experience.

\subsection{Limitations and Future Work}
\rev{This research comes with some limitations which suggests fruitful avenues for future research. First, our measurement of emotions relies on detecting keywords categorized into specific emotional categories, without considering the context in which these keywords are used. Future research should aim to incorporate contextual nuances into emotion detection systems while ensuring transparency in their operations \cite{teodorescu2023language, 17emo}. Second, while this study investigated the effects of emotion self-monitoring and emotion peer-monitoring independently, future work could explore the combined impact of providing individuals with emotional analyses of both their own and their peers' written comments. Such an approach may reveal new insights into collaborative dynamics and emotional awareness. Third, as participants in both treatment groups complained about the transparency of the emotion analysis, future research could provide more transparency in the algorithmic result on the emotion or toxicity score \cite{cscw21-recast}. Future research could focus on enhancing the interpretability of these systems to foster trust and usability \cite{cscw21-recast}. Additionally, we encourage the HCI community to prioritize research on trust calibration with AI systems \cite{trust22} and the development of explainable AI technologies \cite{humanxai21, explain18} to promote responsible and effective AI usage \cite{designai23}. Finally, future work could expand beyond linguistic markers to enhance emotion detection and communication. Exploring multimodal features such as speech \cite{speech22}, facial expressions \cite{emoglass22}, and tactile cues like keystroke strength \cite{keystroke20} could significantly advance the field.}

%% file: tables/discussion.tex
\begin{tabular}{>{\raggedright\arraybackslash}p{6em}>{\raggedright\arraybackslash}p{15em}>{\raggedright\arraybackslash}p{15em}}
\toprule
\textbf{Topic} & \textbf{RQ1: Effects of Emotion Peer-Monitoring} & \textbf{RQ2: Effects of Emotion Self-Monitoring} \\
\toprule
\textbf{Participants reactions} & They thought the analysis was accurate and tried to adjust their wordings to empathy with others. & They thought the analysis was inaccurate and reported the discrepancy between emotional feedback and their true feelings. \\
\midrule
\textbf{Inferred emotions in the message} & Heightened negative emotions, reduced positive emotion and reduced hate speech from female participants, while the reversed trend from male participants. & Heightened negative emotions and a reduced positive emotion from both genders, but only female participants reduced hate speech. \\
\midrule
\textbf{Self-reported emotions by participants} & Participants did not view the peer's emotions differently compared to CG and they viewed their own emotions less negatively. & Participants viewed the peer's emotions more negatively compared to CG and they viewed their own emotions differently for different genders. \\
\midrule
\textbf{Linguistic usage} & Participants used more pronouns and fewer articles, statistical significance was detected in female participants. & Participants used more auxiliary verbs and fewer conjunctions, statistical significance was detected in female participants. \\
\bottomrule
\end{tabular}
\captionof{table}{Summary of findings for RQ1 and RQ2.}
\label{tab:discussion}

%% file: sections/07_conclusion.tex
\section{Conclusion}
Hate speech and toxic communication on social media platforms are on the rise threatening the fabric of society. In addition to reactive moderation, proactive moderation that provides users with real-time, personalized support to engage more constructively in dialogue may be key to further reducing problematic content. Based on the assumptions that discussions on social media platforms are susceptible to lower levels of emotional awareness, we designed and tested an emotion monitoring dashboard to help users perceive the emotions in their own and their peers posts. 

Our study illuminates the intricate interplay between emotion monitoring, linguistic expression, and users' emotional experiences in online environments. The observed increase in negative emotional expression, particularly anger and sadness, suggests that emotion monitoring can heighten self-awareness and potentially lead to more genuine emotional expressions. This aligns with existing research indicating that increased self-awareness of emotions can sometimes exacerbate negative affect, especially when individuals focus intensely on monitoring and regulating their emotional states. The discrepancy between users' self-reported emotions and the emotional content of their messages points to the complexities of emotional expression in digital communication. Users' attempts to moderate their message tone to align with their desired emotional presentation indicate a conscious effort to manage their online persona. However, the challenges in accurately conveying these emotional states underscore the limitations of current emotion detection technologies. Future research should focus on enhancing the accuracy and granularity of emotion detection algorithms to better support users in expressing their true emotions and improving the overall quality of online interactions. By advancing our understanding of how emotion monitoring can shape online conversations, we can develop tools that foster healthier and more empathetic digital communities.

\section*{Ethical Statement}
The experiment was done according to the ethical guidelines of our university. An ethical assessment was done and reviewed according to the internal criteria. An external approval were not required for the study on human participants in accordance with the local legislation and institutional requirements of a Western European university.

%% file: sections/09_appendix.tex
\clearpage
\section{Demographics data of participants to the design workshop}
\label{appendix:demographics}
\begin{center}
\begin{tabular}{p{5em}p{4em}p{6em}p{7em}}
\toprule
\textbf{Participant} & \textbf{Age} & \textbf{Gender} & \textbf{Occupation} \\
\hline
P1 & 34 & Female & UX Research \\
P2 & 32 & Female & UX Designer \\
P3 & 28 & Male & Product Manager \\
P4 & 28 & Male & Product Designer \\
P5 & 26 & Female & Graphic Designer \\
P6 & 28 & Female & Graphic Designer \\
P7 & 27 & Male & UX Researcher \\
P8 & 28 & Female & UX Researcher \\
P9 & 31 & Male & Product Designer \\
P10 & 28 & Female & UX Researcher \\
P11 & 32 & Female & Digital Marketer \\
P12 & 28 & Male & Engineer \\
P13 & 27 & Male & Engineer \\
P14 & 26 & Male & UX Researcher \\
P15 & 28 & Male & Graphic Designer \\
\bottomrule
\end{tabular}
\captionof{table}{Demographics data of fifteen participants to the design workshop for emotion monitoring dashboard.}
\end{center}

\section{Emotional effects: Google T5 analysis}
\label{appendix:t5}
To corroborate our results in the emotional effects study, we also used Google's T5-base model \cite{t520} fine-tuned for emotion recognition \footnote{https://huggingface.co/mrm8488/t5-base-finetuned-emotion} to evaluate the original comments and reframed comments. Instead of detecting the percentage of emotions in each message as \textit{text2emotion} did, this model classifies each message into only one emotion class. However, it is worth noticing that the model contains only four emotions: Angry, Fear, Sad, and Happy. Our findings, illustrated in Figure \ref{fig:preb_t5}, show an increase in the number of messages that express Happy and a decrease in the number of messages that present negative emotions post-reframing. 

\begin{center}
\frame{\includegraphics[width=0.45\columnwidth]{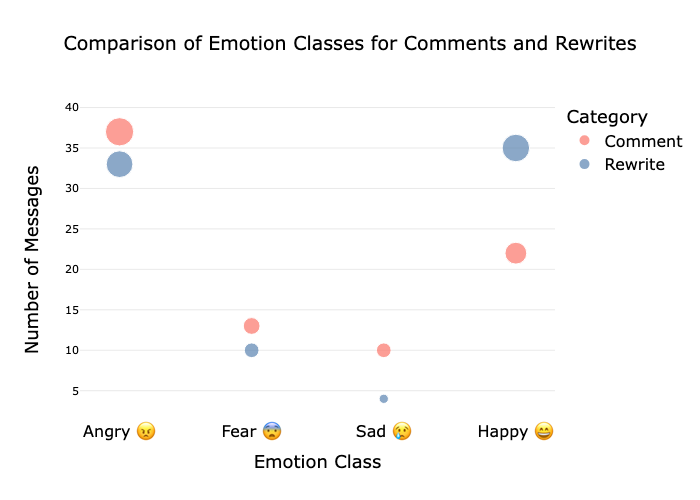}}
\captionof{figure}{Comparison of the number of messages in four emotion classes (Angry, Fear, Sad, Happy) between original and rewritten comments detected by the fine-tuned T5 model from HuggingFace.}
\label{fig:preb_t5}
\end{center}

\newpage
\section{Interface of Emotion Monitoring System.}
\label{appendix:interface}
\begin{center}
\includegraphics[width=0.67\columnwidth]{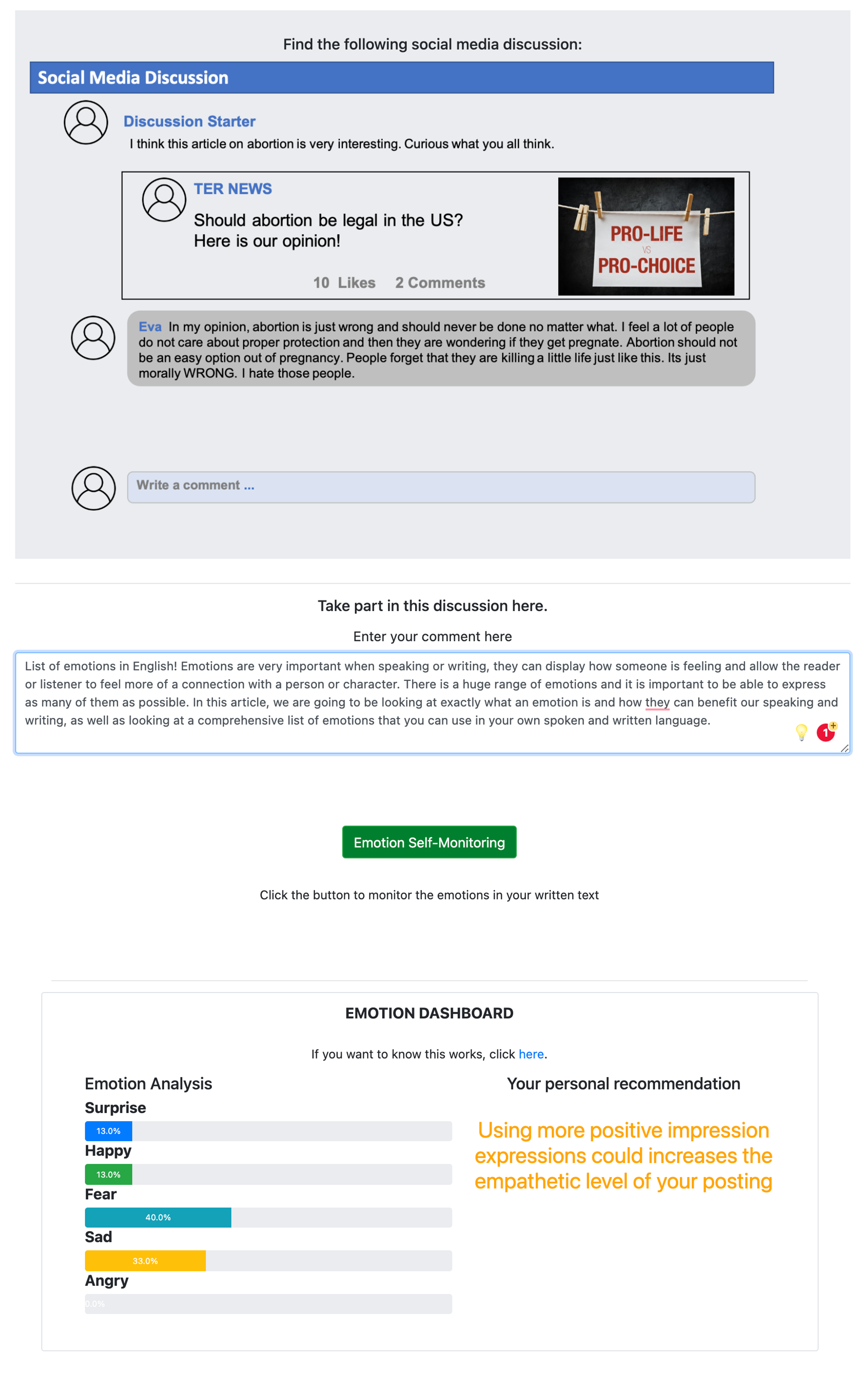}
\captionof{figure}{Interface of Treatment Group 2 where users receive emotion self-monitoring. Participants are presented with a social media discussion containing the news and another user's message, they are asked to write a comment to address the previous user's view. Then they can click the "Emotion Self-Monitoring" button and receive emotional feedback on their own messages. The emotion dashboard presents the final result, it shows emotion analysis on five categories with a personal recommendation.}
\end{center}